\documentstyle[preprint,axodraw,aps]{revtex}
\begin{document}
\title{ {\bf Pion Electromagnetic Current in the Light-Cone Formalism 
\footnote{Sub. Phys.Lett. {\bf B} (1997). }}}
\author{J. P. B. C. de Melo$^a$ , H. W. L. Naus$^b$ and T. Frederico$^c$}
\address{\ \\
$^a$ Instituto de F\'\i sica, Universidade de S\~ao Paulo \\
01498-970 S\~ao Paulo, S\~ao Paulo, Brazil \\
$^b$ Institute for Theoretical Physics, University of Hannover, \\
Appelstr. 2, 30167 Hannover, Germany \\
$^c$ Dep. de F\'\i sica, Instituto Tecnol\'ogico da Aeron\'autica, \\
Centro T\'ecnico Aeroespacial, 12.228-900 S\~ao Jos\'e dos Campos,\\
S\~ao Paulo, Brazil}
\date{\today}
\maketitle

\vspace{2 true cm}

%%\vspace{2 true cm}

\section*{Abstract}

The electromagnetic form factor of the pion is calculated in a pseudoscalar
field theoretical model which constituent quarks. We extract the form factor
using the ``+" component of the electromagnetic current in the light-cone
formalism. For comparison, we also compute the form factor in the covariant
framework and we obtain perfect agreement. It is shown that the pair terms
do not contribute in this pseudoscalar model. This explains why a naive
light-cone calculation, {\it i.e.}, omitting pair terms from the onset, also
yields the same results.

\newpage

\section{Introduction}

The pion, as quark-antiquark bound state, is an appropriate system to study
aspects of QCD in low and intermediate energies regions. In the
nonperturbative regime of QCD, the pion has indeed received much attention,
e.g. using the constituent quark model in the light-cone formalism \cite
{ji90,coest88,inna94}. In these studies, where the pion is described by
light-cone wave functions, the electromagnetic form factor has been
calculated for low and high $q^2$ and a fairly good agreement with
experiment has been obtained.

\begin{figure}[h]
\begin{center}
\centerline{\ 
\begin{picture}(330,200)(0,0)
\Line(0,52)(60,52)
\Line(0,48)(60,48)
%\Line(10,54)(15,50)
%\Line(10,46)(15,50)
\put(30,30){\makebox(0,0)[br]{$P$}}
\put(230,30){\makebox(0,0)[br]{$P^{'}$}}
\Vertex(60,50){3.0}
\ArrowLine(60,50)(200,50)
\Vertex(200,50){3.0}
\ArrowLine(60,50)(130,140)
\ArrowLine(130,140)(200,50)
\Photon(130,140)(130,200){3}{8.5}
\Vertex(130,140){3.0}
\put(220,120){\makebox(0,0)[br]{$P'- k$}}
\put(70,120){\makebox(0,0)[br]{$P - k$}}
\put(130,30){\makebox(0,0)[br]{$k$}}
\Line(200,48)(260,48)
\Line(200,52)(260,52)
%\Line(225,56)(230,50)
%\Line(225,46)(230,50)
\end{picture}
}
\end{center}
\caption{ Feynman triangle diagram with the corresponding momenta.
We use $P$ for the initial state and $P^{\prime}
$ for final state of the pion, $q=P^{\prime}-P$ is the momentum transfer.}
\label{fig1}
\end{figure}
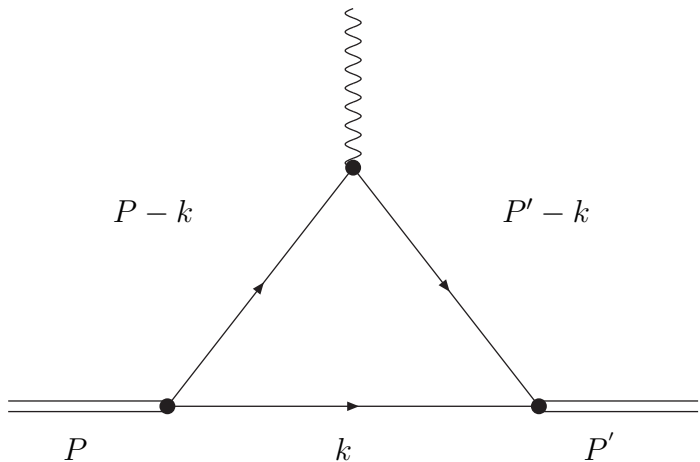

In general, the null-plane is invariant under kinematical transformations
(see \cite{gibi} and references therein). The description in term of wave
functions on the null-plane, however, violates covariance \cite{pa97}. In a
recent work \cite{tob92}, the structure of the pion is formulated in terms
of the triangle diagram (Fig.\ref{fig1}). By means of the Feynman approach
on the light-cone the pion form factor and charge radius have been
calculated. For the actual extraction of these observables the ``+"
component of the electromagnetic current has been utilized. In this case,
pseudoscalar particles and coupling, full covariance is respected.

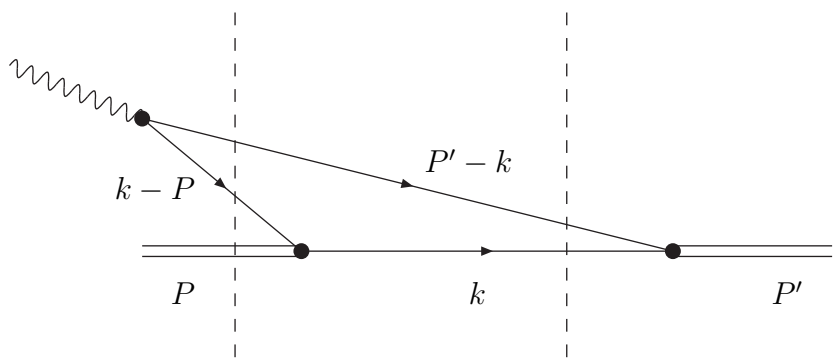
\begin{figure}[h]
\begin{center}
\centerline{\ 
\begin{picture}(330,130)(0,0)
\Line(0,52)(60,52)
\Line(0,48)(60,48)
%\Line(10,54)(15,50)
%\Line(10,46)(15,50)
\Vertex(60,50){3.0}
\ArrowLine(0,100)(60,50)
\ArrowLine(60,50)(200,50)
\Vertex(200,50){3.0}
%\ArrowLine(200,50)(0,100)
\ArrowLine(0,100)(200,50)
\Line(200,48)(260,48)
\Line(200,52)(260,52)
%\Line(220,54)(230,50)
%\Line(220,46)(230,50)
%\Photon(-50,120)(0,100){4}{8.5}
\Photon(-50,120)(0,100){3}{8.5}
\Vertex(0,100){3.0}
\put(20,70){\makebox(0,0)[br]{$k-P$}}
\put(20,30){\makebox(0,0)[br]{$P$}}
\put(130,30){\makebox(0,0)[br]{$k$}}
\put(140,80){\makebox(0,0)[br]{$P'-k$}}
\put(250,30){\makebox(0,0)[br]{$P'$}}
\DashLine(35,140)(35,10){5}
\DashLine(160,140)(160,10){5}
\end{picture}
}
\end{center}
\caption{Pair creation diagram}
\label{fig2}
\end{figure}

In several light-cone studies \cite{sa,pasu} the role of the so-called pair
terms, {\it i.e.}, particle--antiparticle pair creation by the photon (Fig.%
\ref{fig2}), has been extensively discussed. In ref. \cite{pasu}, we studied
pair terms in an explicit computation of the electromagnetic current of
scalar and vector mesons. It was shown that these pair terms are essential
for retaining full covariance. Recently \cite{Nau}, we also demonstrated in
a boson model the relevance of pair terms for the Ward-Takahashi identity.
The latter expresses local gauge invariance for (off-shell) Greens functions.

In this work we calculate the pion form factor in a similar light-cone model
as in \cite{tob92}, however carefully adressing the issue of pair terms. We
restrict ourselves to the study of on-shell current matrix elements. It is
shown that pair term contributions eventually disappear. We argue, however,
that this is a coincidence and in general not true. In other words, this
cancellation presumably is a property of the pseudoscalar model under
consideration.

\section{Matrix Elements}

As in earlier applications \cite{tob92}, we use an effective Lagrangian
approach with pion and quark degrees of freedom. We choose pseudoscalar
coupling: 
\begin{equation}
{\it L_I= - \imath \frac{m}{f_\pi} \vec\pi \cdot \overline q \gamma^5 \vec 
\tau q \, ,}  \label{lain}
\end{equation}
where $m$ denotes the consituent quark mass and $f_\pi$ the pion decay
constant. The electromagnetic field is coupled in the usual minimal way,
ensuring gauge invariance. The light-cone coordinates are defined as $%
k^+=k^0+k^z \ , k^-=k^0-k^z \ , k_\perp=(k^x,k^y)\, .$ For the ``+"
component of the electromagnetic current of the $\pi^+$, we get the
expression corresponding to the Feynman triangle diagram (Fig.\ref{fig1}): 
\begin{eqnarray}
J^+ &=&\imath 2 e \frac{m^2}{f^2_\pi} N_c\int \frac{d^4k}{(2\pi)^4} Tr[S(k)
\gamma^5 S(k-P^{\prime}) \gamma^+ S(k-P) \gamma^5 \Lambda(k,P^{\prime})
\Lambda(k,P) ] \ ,  \label{j+pion2}
\end{eqnarray}
with $\displaystyle S(p)=\frac{1}{\rlap\slash p-m+\imath \epsilon} \, .$ $%
N_c=3$ is the number of colors and the factor 2 stems from isospin algebra.
We will work in the Breit-frame, where $P^0=P^{\prime \, 0}$ and $\vec P
^{\prime}_\perp=-\vec P_\perp=\vec \frac{q_\perp}{2}$. The
function %\be
$\displaystyle \Lambda(k,p)=\frac{N}{((p-k)^2-m^2_{R}+\imath \epsilon)}$ 
%\ee
is our choice for regularizing the divergent integral. 
%Feynman triangle diagram.
The normalization constant $N$ is found by imposing the condition $%
F_{\pi}(0)=1$ on the pion form factor. 
%After integration in $k^-$, this function, expressed in light-cone coordinates, is interpretated as the pion wave function \cite{pa97,tob92,shakin}.

The trace in light-cone coordinates is given by 
\begin{equation}
Tr_\pi =-4 k^- (k^+-P^+)^2 + 4 (k^2_\perp+m^2) (k^+-2 P^+) + k^+ q^2 \ .
\end{equation}
Rewriting Eq.(\ref{j+pion2}) in light-cone coordinates yields 
\begin{eqnarray}
J^{+}&=& 2 \imath e \frac{m^2 N^2}{f^2_\pi} N_c \int \frac{d^{2} k_{\perp} d
k^{+} d k^-}{2(2 \pi)^4} \frac{-4 k^- (k^+-P^+)^2 + 4 (k^2_\perp+m^2) (k^+-2
P^+) + k^+ q^2 } {k^+(P^+-k^+)^2 (P^{^{\prime}+}-k^+)^2 (k^- - \frac{%
f_1-\imath \epsilon }{k^+})} \\
& &\frac{1} {(P^- - k^- - \frac{f_2 -\imath \epsilon }{P^+ - k^+})
(P^{\prime -} - k^- - \frac{f_3 -\imath \epsilon }{P^{\prime +} - k^+}) (P^-
- k^- - \frac{f_4 -\imath \epsilon }{P^+ - k^+}) (P^{\prime -} - k^- - \frac{%
f_5 -\imath \epsilon }{P^{\prime +} - k^+})} \ ,  \nonumber
\end{eqnarray}
where $f_1=k_{\perp}^{2}+m^2$ , \ \ $f_2=(P-k)_{\perp}^{2}+m^2$ , \\$%
f_3=(P^{\prime}-k)_{\perp}^{2}+m^2$ , \ \ $f_4=(P-k)_{\perp}^{2}+m^2_{R}$ \
and $f_5=(P^{\prime}-k)_{\perp}^{2}+m^2_{R}$ \ .

According to ref. \cite{pasu}, the ``bad terms" $(Tr_\pi^B)$ in the matrix
element are terms proportional to $k^-$, the light-cone energy in the loop
integral and the ``good terms" depend on $k_\perp$ and $k^+$. Therefore we
define the traces 
\begin{eqnarray}
Tr_\pi^G & = &4 (k^2_\perp+m^2) (k^+-2 P^+) + k^+ q^2 \; , \\
Tr_\pi^B & = & -4 k^- (k^+-P^+)^2 \, .  \nonumber
\end{eqnarray}

For the good terms, only momenta in the interval $0<k^+<P^+$ contribute to
the Cauchy integration; this means that the spectator particle is on
mass-shell and the pole contribution is $k^-=(k_\perp+m^2)/k^+$ . We also
construct the bad terms for the pion current matrix elements: 
\begin{eqnarray}
\Delta_{\pi}&=& \int \frac{d^2 k_{\perp} d k^{+} d k^-}{2(2 \pi)^4} \frac{
k^- (k^+-P^+)^2 } {k^+(P^+-k^+)^2 (P^{{\prime}+}-k^+)^2 (k^- - \frac{%
f_1-\imath \epsilon }{k^+})}  \label{bterm} \\
& & \frac{1} {(P^- - k^- - \frac{f_2 -\imath \epsilon }{P^+ - k^+})
(P^{\prime -} - k^- - \frac{f_3 -\imath \epsilon }{P^{\prime +} - k^+}) (P^-
- k^- - \frac{f_4 -\imath \epsilon }{P^+ - k^+}) (P^{\prime -} - k^- - \frac{%
f_5 -\imath \epsilon }{P^{\prime +} - k^+})} \ .  \nonumber
\end{eqnarray}
This integral has contributions in two nonzero intervals:

I) $0<k^+ < P^+$ and

II) $P^+ < k^+ < P^{\prime +}$ , where $P^{\prime +}=P^+ + \delta$ \ .

Note that in the Breit frame $P^+=P^{\prime +}$, which implies that there
appear coinciding poles in Eq.(\ref{bterm}). 
%that the first two poles in Eq.(\ref{bterm}) coincide.
As in \cite{Nau}, we have dislocated them by shifting $P^{\prime +}$ with $%
\delta$. The interval (I) corresponds to a spectator particle on mass-shell.
The other interval (II) corresponds to a pair term contribution (Fig. \ref
{fig2}). Eventually, we take the limit $\delta \rightarrow 0 $, {\it i.e.}, $%
P^{\prime +}=P^+$, and the exact kinematics of the Breit frame is recovered.

Let us consider interval II, $P^{+}<k^{+}<P^{\prime }{}^{+}$; after
integration in $k^{-}$ we obtain 
\begin{eqnarray}
\Delta _\pi ^{II} &=&\imath \int \frac{d^2k_{\perp }dk^{+}}{2(2\pi )^3}\frac{%
(P^{\prime }{}^{-}-\frac{f_3}{P^{\prime }{}^{+}-k^{+}})}{k^{+}(P^{^{\prime
}+}-k^{+})^2(P^{\prime -}-\frac{f3}{P^{^{\prime }+}-k^{+}}-\frac{f_1}{k^{+}})%
} \\
&&\frac{\theta (P^{\prime }{}^{+}-k^{+})\theta (k^{+}-P^{+})}{(\frac{f_3}{%
P^{^{\prime }+}-k^{+}}-\frac{f_2}{P^{+}-k^{+}})(\frac{f_3}{P^{^{\prime
}+}-k^{+}}-\frac{f_4}{P^{+}-k^{+}})(\frac{f_3}{P^{^{\prime }+}-k^{+}}-\frac{%
f_5}{P^{+}-k^{+}})}\ .  \nonumber
\end{eqnarray}
The limit to the Breit frame is performed, after the
momentum fraction is used as integration variable; $x=(k^{+}-P^{+})/(P^{%
\prime +}-P^{+})$ . In the limit of $\delta \rightarrow 0$ the integration
becomes:

\begin{eqnarray}
\Delta _\pi ^{II} &=&\imath \frac \delta {P^{+}}\int \frac{d^2k_{\perp }dx}{%
2(2\pi )^3}\frac{\theta (x)\theta (1-x)}{(1-x)^2(\frac{f3}{1-x}+\frac{f_2}x)(%
\frac{f3}{1-x}+\frac{f_4}x)(\frac{f3}{1-x}+\frac{f_5}x)}\rightarrow 0\ ; \\
&&  \nonumber
\end{eqnarray}
which vanishes linearly with $\delta $ when the Breit frame is recovered.
Thus we see that in this model, pseudoscalar coupling between the
constituent fermions and the pseudoscalar pion, the pair terms in the
electromagnetic current disappear. Rather than generalizing this result to
other models in the light-cone formalism, we consider it to be a coincidence
for this particular case. We therefore get contributions of the good terms
and only of the interval I $(\Delta _\pi ^I)$ . As a consequence, one
obtains agreement between naive light-cone and covariant calculations.

\section{Form Factor}

In general the form factor is extracted from the covariant expression: 
\begin{equation}
J^\mu = e (P^{\mu}+P^{\prime \mu}) F_\pi(q^2) \, .
\end{equation}
We only use the ``+" component of the current. The equation for the form
factor written in light-cone coordinates for this model is

\begin{eqnarray}
F_{\pi}(q^2) &=& 2 \imath e \frac{m^2 N^2}{ 2 P^+ f^2_\pi} N_c \int \frac{%
d^{2} k_{\perp} d k^{+} d k^-}{2(2 \pi)^4} \frac{-4 k^- (k^+-P^+)^2 + 4
(k^2_\perp+m^2) (k^+-2 P^+) + k^+ q^2 } {k^+(P^+-k^+)^2
(P^{^{\prime}+}-k^+)^2 (k^- - \frac{f_1-\imath \epsilon }{k^+})} \\
& &\frac{1} {(P^- - k^- - \frac{f_2 -\imath \epsilon }{P^+ - k^+})
(P^{\prime -} - k^- - \frac{f_3 -\imath \epsilon }{P^{\prime +} - k^+}) (P^-
- k^- - \frac{f_4 -\imath \epsilon }{P^+ - k^+}) (P^{\prime +} - k^- - \frac{%
f_5 -\imath \epsilon }{P^{\prime +} - k^+})} \, .  \nonumber
\end{eqnarray}

\begin{figure}[h]
\vspace{15.0cm}
\caption{ Pion form factor as a function of $Q^2 = - q^2$.}
\label{fig3} \includegraphics{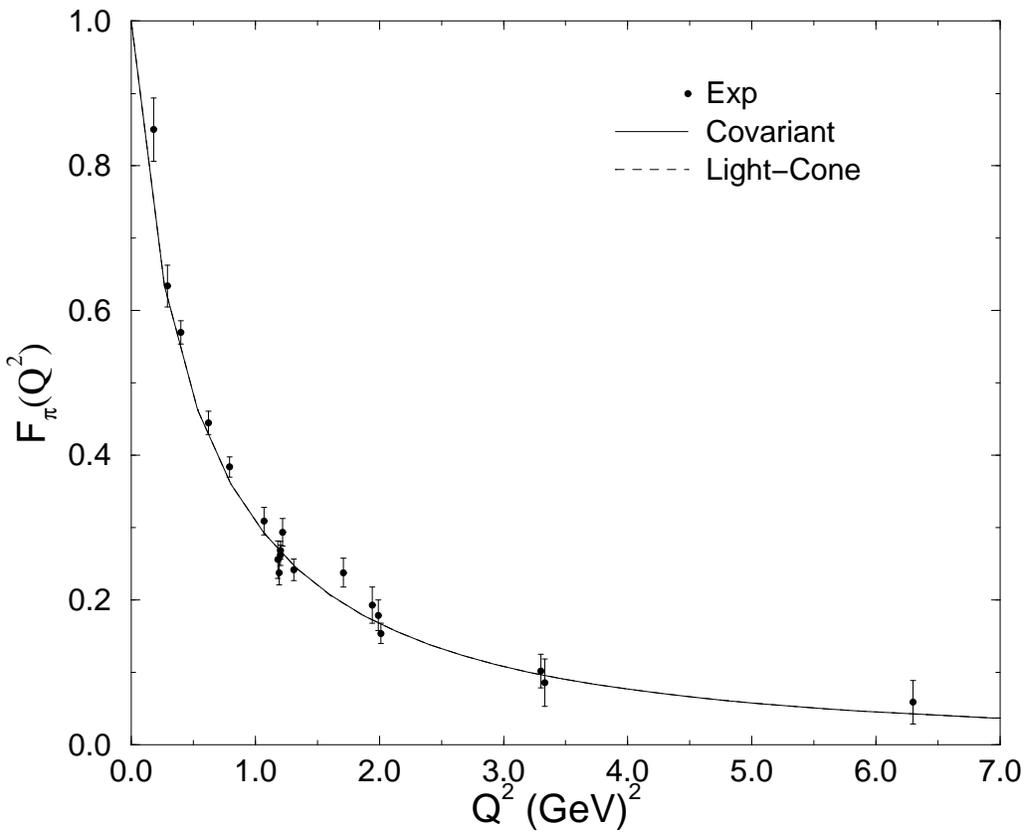} %meson mass  = $m_{\pi}$ = 0.140 GeV, 
%quark and anti-quark mass $m_q$= 0.220 GeV, \\
%regulator mass = $m_{R}$= 0.946 GeV, $f_{\pi}$=0.093 GeV , \\ 
%Exp. - L.J. Bebek et Al. PR{\bf D13}(1978) 1693.}
\end{figure}

One can verify that only the on-shell pole $k^-=(k_\perp+m^2)/k^+$
contributes to the $k^-$ integration: 
\begin{eqnarray}
F_{\pi}(q^2) &=& \frac{m^2 N^2}{ P^+ f^2_\pi} N_c \int \frac{d^{2} k_{\perp}
d k^{+}}{2(2 \pi)^3} \frac{-4 (\frac{f_1}{k^+})(k^+-P^+)^2 + 4
(k^2_\perp+m^2) (k^+-2 P^+) + k^+ q^2 } {k^+(P^+-k^+)^2
(P^{^{\prime}+}-k^+)^2 } \\
& &\frac{\theta(p^{\prime +}-k^+) \theta(k^+-P^+) }{(P^- - \frac{f_1}{k^+}
- \frac{f_2 }{P^+ - k^+}) (P^{\prime -} - \frac{f_1}{k^+} - \frac{f_3 }{%
P^{\prime +} - k^+}) (P^- - \frac{f_1}{k^+} - \frac{f_4 }{P^+ - k^+})
(P^{\prime +} - \frac{f_1}{k^+} - \frac{f_5 }{P^{\prime +} - k^+})} \, . 
\nonumber
\end{eqnarray}
In this way the null-plane (light-cone) wave function for the $\pi$ meson
appears \cite{pa97,tob92,shakin},

\begin{eqnarray}
\Phi_i(x,k_\perp)=\frac{1}{(1-x)^2} \frac{N}{(m^2_\pi-M^2_{0})
(m^2_{\pi}-M^2_{R})} \ ,
\end{eqnarray}
where $x=k^+/P^+$. $M^2_{R}$ is given by the function 
\begin{equation}
M^2_{R}={\cal M}^2(m^2, m_R^2)= \frac{k^2_\perp+m^2}{x}+\frac{%
(P-k)^2_\perp+m^2_{R}}{1-x}-P^2_\perp \ .
\end{equation}
Recall the free quark-antiquark mass squared: $M^2_0 ={\cal M}^2(m^2, m^2)$.
The form factor is finally written as: 
\begin{eqnarray}
F_{\pi}(q^2)&=& \frac{m^2}{ P^+ f^2_\pi} N_c \int \frac{d^{2} k_{\perp} d x}{%
2(2 \pi)^3} \frac{-4 (\frac{f_1}{x P^+})(x P^+ - P^+)^2 + 4 (k^2_\perp+m^2)
(x P^+-2 P^+) + k^+ q^2 } {x } \nonumber \\
& & \theta(x) \theta(1-x) \Phi^*_f(x,k_{\perp}) \Phi_i(x,k_{\perp})\, . 
\label{form} 
\end{eqnarray}
The remaining integrals are evaluated numerically and the result is
presented in Fig.\ref{fig3}. The two free parameters in this model, the
constituent quark mass $m_q$ and the regulator mass $m_{R}$ were fixed as: $%
m_q= 0.220$ GeV, $m_{R}=0.946$ GeV; for the pion mass we take $m_\pi = 0.140$
GeV. This model form factor calculated in the light-cone framework agrees
with the one obtained in the covariant formalism (see also Fig.\ref{fig3}).
In the covariant calculation the energy integral, {\it i.e.}, the $k^0$
integral, is obtained analytically via Cauchy's theorem. Again, the
remaining part is computed numerically. Furthermore, in Fig.\ref{fig3}, we
compare the calculated model pion form factors to experimental data \cite
{bebek} and find good agreement. %The contribution of the pair term in
%the bad matrix elements vanishes, cf. section II. 
%Thus, the remaining terms --good terms
%and the on mass-shell spectator bad term-- exactly
%yield  the covariant result.

\section{Pion Decay Constant}

The pion decay constant $f_\pi$ is measured in the weak leptonic decay of
the charged pion and appears in the matrix element of the partially
conserved axial vector current: 
\begin{equation}
P_\mu <0|A^\mu_i |\pi_j>= \imath m_\pi^2 f_\pi \delta_{ij} \, .
\end{equation}
%In the partial conservation of axial-vector current (PCAC), the 
%constant weak decay ($f_\pi$) is expressed by 
%matriz elements \cite{tob92}:
Following ref.\cite{tob92}, we take $A^\mu_i = \bar{q} \gamma^\mu \gamma^5 
\frac{\tau_i}{2} q$ and use the interaction Lagrangian Eq.(\ref{lain}) for
the pion-$\bar{q} q$ vertex function. In this way we obtain %\be
%p_\mu <0|\bar{q} \gamma^\mu \gamma^5 \tau_i q|\pi_j>=2 \imath p^2 
%f_\pi \delta_{ij} \ ,
%\ee
%
%wich governs the leptonic decay of the charged pion via the axial 
%current.
% 
%In this model, we have :

\begin{eqnarray}
\imath P^2 f_\pi &=& \frac{m}{f_\pi} N_c\int \frac{d^4k}{(2\pi)^4} Tr[%
\rlap\slash P \gamma^5 S(k) \gamma^5 S(k-P) \Lambda(k,P) ] \ .  \label{f_pi}
\end{eqnarray}
After calculating the trace in the pion center of mass system, this equation
is rewritten in light-cone coordinates as

\begin{equation}
\imath m_\pi^2 f_{\pi} = N \frac{m}{f_{\pi}} N_c \int \frac{d^{2} k_{\perp}
d k^{+} d k^-}{2 \ (2 \pi)^4 } \frac{ -4 \ m \ m_{\pi}^2 } {k^+(P^+-k^+)
(k^- - \frac{f_1-\imath \epsilon }{k^+}) (P^- - k^- - \frac{f_2 -\imath
\epsilon }{P^+ - k^+}) (P^- - k^- - \frac{f_4 -\imath \epsilon }{P^+ - k^+})}
\ .
\end{equation}
%
%Where $N$ is the normalization constante wave function.
%
Perfoming the Cauchy integration in $k^-$ results in:

\begin{equation}
f_{\pi}^{2} = N N_c \int \frac{d^{2} k_{\perp} d k^{+}} { (2 \pi)^3} \frac{
2 m^2 } { k^+(P^+-k^+) (P^- - \frac{f_1}{P^+} - \frac{f_2}{P^+ - k^+}) (P^-
- \frac{f_1}{P^+} - \frac{f_4}{P^+ - k^+})} \ .
\end{equation}

In terms of the model light-cone wave function we get as final expression
for the weak decay constant $f_{\pi}$ :

\begin{equation}
f_{\pi}^{2} = N_c \int \frac{d^{2} k_{\perp} d x } {(2 \pi)^3} \frac{2 \ m^2 
} {x} \Phi(x,k_\perp) \ .
\end{equation}
Numerically, this yields $f_\pi = 101$ MeV to be compared with the
experimental value $f_\pi = 93$ MeV. Similar discrepancies were found in
refs.\cite{ji90,tob92} and appear to be a property of these models.

\section{Summary}

%In this model, we have two free parameters, 
%the constituent quark mass 
%($m_q$ $=$ $m_{\bar q}$$=$ $0.220$ GeV) and  the
%regulator mass ($m_{R}$=$0.946$ GeV). 
%For the pion decay constant we take the experimental
%value $f_{\pi}$=$93.0$ MeV.

In a pseudoscalar constituent quark model, we calculated the pion form
factor in light-cone as well as covariant field theory. The results are in
perfect agreement with each other and also describe the experimental form
factor well in the $q^2$--range considered. In the light-cone formalism we
have explicitly shown that the contribution of the pair terms in the bad
matrix elements vanishes. Thus, the remaining terms --good terms and the on
mass-shell spectator bad term-- exactly yield the covariant result. Since
this is a peculiar property of the model under consideration, this does not
justify to omit the pair terms in general. It merely explains the success of
such a naive light-cone computation in this particular case.

\section*{Acknowledgments}
This work was supported in part by
Funda\c c\~ao Coordena\c c\~ao 
de Aperfei\c coamento de Pessoal de N\'\i vel Superior and 
the Deutscher Akademischer Austauschdienst
(Probral/CAPES/DAAD project 015/95).
It was also supported by 
the Brazilian agencies CNPq and FAPESP. 
J. P. B. C. M. acknowledges the hospitality of the
Institute for Theoretical Physics, University of Hannover.

%%%%%\end{document}

\newpage

%%%%%%%%%% figure by Axodraw %%%%%%%%%%%%%%%%%%%%%%%%%%

\vskip 2cm

\newpage 
.
\vskip 2cm

\end{document}